\documentstyle[12pt,epsfig,multirow]{article}

\newcommand{\be}{\begin{eqnarray}}
\newcommand{\ee}{\end{eqnarray}}
\newcommand{\etal}{{\it et al.}}
\def\nue{{\nu_e}}
\def\anue{{\bar\nu_e}}
\def\numu{{\nu_{\mu}}}

\newcommand{\ms}{\Delta m^2_{21}}
\newcommand{\ma}{\Delta m^2_{31}}

\newcommand{\sss}{\sin^2 \theta_{12}}
\newcommand{\sch}{\sin^2 \theta_{13}}

\newcommand{\ses}{\sin^2 \theta_{14}}
\newcommand{\sess}{\sin^2 \theta_{15}}
\newcommand{\peebar}{P_{\bar e \bar e}} 

\def\ltap{\ \raisebox{-.4ex}{\rlap{$\sim$}} \raisebox{.4ex}{$<$}\ }

\begin{document}

\begin{flushright}
\texttt{HRI-P-07-07-003}\\
\end{flushright}
\bigskip

\begin{center}
{\Large \bf The (3+2) Neutrino Mass Spectrum and Double Chooz}

\vspace{.5in}

{\bf Abhijit Bandyopadhyay$^{a}$ and  
Sandhya Choubey$^{b}$}
\vskip .5cm
{\normalsize \it Harish-Chandra Research Institute,} \\
{\normalsize \it Chhatnag Road, Jhunsi, Allahabad  211019, India}
\vskip 1cm
\vskip 2cm

{\bf ABSTRACT}
\end{center}
The implications of extra sterile neutrinos 
for the Double Chooz experiment is expounded.
The so-called ``3+2'' mass  
spectrum with 2 sterile neutrinos mixed with the active ones,
is still allowed by the global neutrino data including 
MiniBooNE. 
We probe its impact on the resultant 
reactor antineutrino signal at the near and far detector
of the Double Chooz experiment. The oscillations 
driven by the additional mass squared difference due to 
the sterile states bring an energy independent constant 
suppression at both the near and far detectors. 
We study to what extent the measurement of 
$\theta_{13}$ would get affected due to the presence of 
sterile mixing. We also give the projected sensitivity that 
Double Chooz will have to constrain the extra mixing angles 
associated with the sterile states.

\vskip 5cm

\noindent $^a$ email: abhi@mri.ernet.in 

\noindent $^b$ email: sandhya@mri.ernet.in

\newpage

\section{Introduction}

Determining the mixing angle $\theta_{13}$ is the next 
priority in the field of neutrino oscillation physics. 
Discovery of a non-zero value for this mixing angle 
is a prerequisite for determining the two other 
ingredients of the 
neutrino mass matrix\footnote{The remaining 
unknowns, which include the absolute neutrino mass 
scale and the Majorana phases (if neutrinos are indeed Majorana 
particles) cannot be ascertained in neutrino oscillation 
experiments.}, 
{\it viz}, the 
CP phase $\delta_{CP}$ and the sign of the 
atmospheric neutrino mass squared difference $sgn(\ma)$.
Various 
experimental proposals have been put forward to 
resolve this perplexing issue and measure this 
hitherto unknown mixing angle. Accelerator based experiments 
such as T2K and NO$\nu$A \cite{t2knova}
involving conventional neutrino beams from pion decays 
are expected to come up in the near future. Superbeam 
upgrades of these facilities are also being 
envisaged. 
Pure $\nue$ and/or $\anue$ fluxes from beta decay of 
highly accelerated radioactive ions stored in rings 
is called Betabeam and under ingenious 
experimental set-ups can give extremely good sensitivity to 
$\theta_{13}$ \cite{betabeam}. 
The ultimate neutrino oscillation machine of 
course would be the neutrino factory \cite{nufact}, 
which could provide 
unprecedented $\theta_{13}$ sensitivity, if indeed this 
mixing angle turns out to be extremely small. 
The 
main neutrino oscillation channel probed in these experiments 
are either the $\numu\rightarrow \nue$ conversion channel $P_{\mu e}$
or the $\nue\rightarrow \numu$ conversion channel $P_{e \mu}$, 
which is popularly known as the ``Golden Channel''. 
The conversion channels suffer from the intrinsic 
problem of parameter degeneracies, whereby 
one ends up with multiple fake solutions in addition to the true one.
The three kinds of parameters degeneracies 
are the so-called 
($\theta_{13},\delta_{CP}$) intrinsic degeneracy 
\cite{intrinsic},
the ($sgn(\ma),\delta_{CP}$) degeneracy \cite{minadeg}, and 
the ($\theta_{23},\pi/2-\theta_{23}$) degeneracy 
\cite{th23octant}, and they together 
lead to a
total eight-fold degeneracy \cite{eight} of parameter values.

The oscillation channel completely free of parameter degeneracies 
is the $\nue\rightarrow \nue$ survival channel, $P_{ee}$. 
This channel 
can be effectively probed in reactor based experiments. 
The crucial issue which needs to be addressed for maximum 
$\theta_{13}$ sensitivity is reducing the systematic uncertainties.
It has been widely accepted that the best way of achieving this 
is by performing the experiment with two (or more) detector 
set-up. In these experiments a ``near'' detector is placed 
very close to the reactor cores and another one 
farther away, at the baseline optimal for observing near-maximal 
oscillations driven by $\ma$ \cite{white}. 
Double Chooz \cite{chooz2letter,chooz2} is one 
such up-coming experiment 
which proposes to use the Chooz-B 
nuclear power plant in France. 
The other proposals that are being considered 
include ANGRA in Brazil \cite{angra},
Daya Bay in China \cite{dayabay},
and RENO in South Korea \cite{reno}.
The first idea for using a near-far detector set-up 
in reactor experiments for measuring $\theta_{13}$ 
was put forth by the KR2DET collaboration in Russia \cite{kr2det}.
However, this experiment was shelved due to various reasons.
Two other very good proposals which were eventually turned 
down by funding agencies include the experimental proposal 
to use the 
Braidwood reactor in the U.S.A. \cite{braidwood} and  
KASKA in Japan \cite{kaska}.

There are two-fold advantage of 
the experimental set-up with reactor 
neutrinos. 
Firstly as mentioned 
above, these experiments use the $P_{ee}$ channel, which 
is free from the problems of parameter degeneracies.
Secondly, there are no problems of matter effects 
\cite{msw1,msw2,magicfirst} in these 
experiments. However, there are at least two other experimental 
scenarios where very large matter effects in $P_{ee}$
can be exploited to provide stringent tests of $\theta_{13}$ 
as well as $sgn(\ma)$. It has been realized that 
extremely large matter effects inside the supernova 
leaves an imprint on the resultant neutrino spectrum 
and hence 
in principle its possible to  
determine $\theta_{13}$ and $sgn(\ma)$ 
from the neutrino signal of a 
future galactic supernova \cite{sn,sn3p2}.
Very recently there has been a suggestion that in very long 
baseline experiments involving pure $\nue/\anue$ 
fluxes such as Betabeams, one could effectively use the 
very large matter effects in $P_{ee}$ to pin down $\theta_{13}$ and 
$sgn(\ma)$ \cite{pee}. 

The recent declaration of the MiniBooNE results \cite{miniboone}
has brought back the issue of sterile neutrinos on the 
front line of neutrino physics. The MiniBooNE 
experiment was designed to test the oscillation 
claim of the LSND experiment \cite{lsnd}.
Since the 
oscillation interpretation of the LSND signal demands a 
mass squared difference $\Delta m^2 \sim $ eV$^2$, it cannot 
be accommodated along with the solar and atmospheric data 
within a three flavor framework. One therefore needs 
one or more extra neutrinos and these species necessarily 
must be sterile. For one extra sterile neutrino 
one would get the so-called 2+2 and 3+1 schemes \cite{sterileold}.
Even before the MiniBooNE results, 
the 2+2 scheme was already disfavored from the solar and 
atmospheric data, while the 3+1 scheme suffered from 
severe tension between the LSND signal and the null 
results observed at other short baseline experiments \cite{sbl}.
However, all data could be fitted if one allows for 2 sterile 
neutrinos mixed with the active ones, leading to the so-called 
3+2 mass schemes \cite {sn3p2,threeplustwo,ws3p2}. 
MiniBooNE has reported null 
signal in the energy range where it would have expected to see 
an excess of electron events if LSND experiment were due to 
two flavor oscillations. Though on the face of it this may 
look like a strong signal against the hypothesis of sterile 
neutrinos, in reality there is still ample room for their 
existence. For neutrino mass spectra with sterile neutrinos 
which predict $\Delta m^2 \sim $ eV$^2$, one can see 
from the detailed analysis presented in \cite{maltonischwetz}
that even though the tension between the positive signal at 
LSND and negative signal at all other short baseline experiment 
including MiniBooNE, rules out the  
3+1 scheme,
the 3+2 scheme with 2 sterile neutrinos is 
still viable if one allows for CP violating phases. 
One should also keep in mind that even if it was proved that 
LSND was indeed wrong, it would only mean that sterile 
neutrinos in the eV$^2$ mass regime are ruled out. 
One could still envisage heavier sterile neutrinos 
which would have implications for astrophysics and 
cosmology, and these neutrinos could be mixed with 
active neutrino species. 

In this paper we look at the implications of these extra 
sterile neutrinos for the Double Chooz experiment. For 
concreteness we work within the allowed framework of the 3+2 
scheme which gives a viable explanation of current world 
neutrino data. 
However, our results can be easily extended to other
scenarios since the oscillation driven by the extra $\Delta m^2$ 
corresponding to the sterile states anyway average out for the 
Double Chooz experiment. 
Our results can also be easily extended to the other 
reactor experiments mentioned before. 
We begin by discussing the oscillation 
probability in the 3+2 scheme in section 2. In section 3 
we present the expected events in the near and far detector 
of Double Chooz if the 3+2 scheme was correct. Section 4
has our results with the full statistical analysis of the 
projected data set of Double Chooz. We end in section 5
with discussions and conclusions.

\section{Oscillation Probability in the 3+2 Scheme}

\begin{figure}[t]
\includegraphics[width=8.0cm, height=7cm, angle=0]{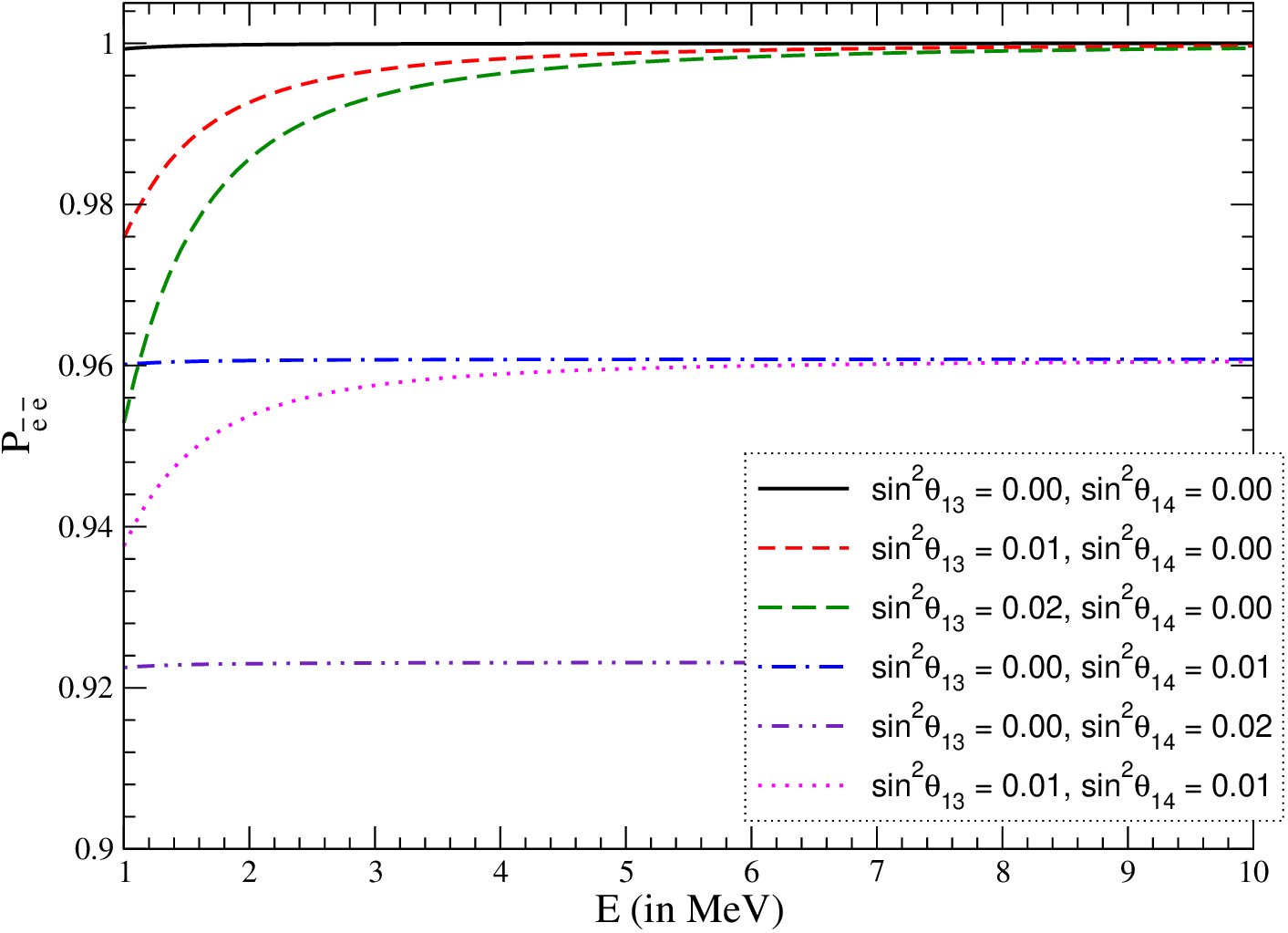}
\vglue -7.0cm \hglue 9.0cm
\includegraphics[width=8.0cm, height=7cm, angle=0]{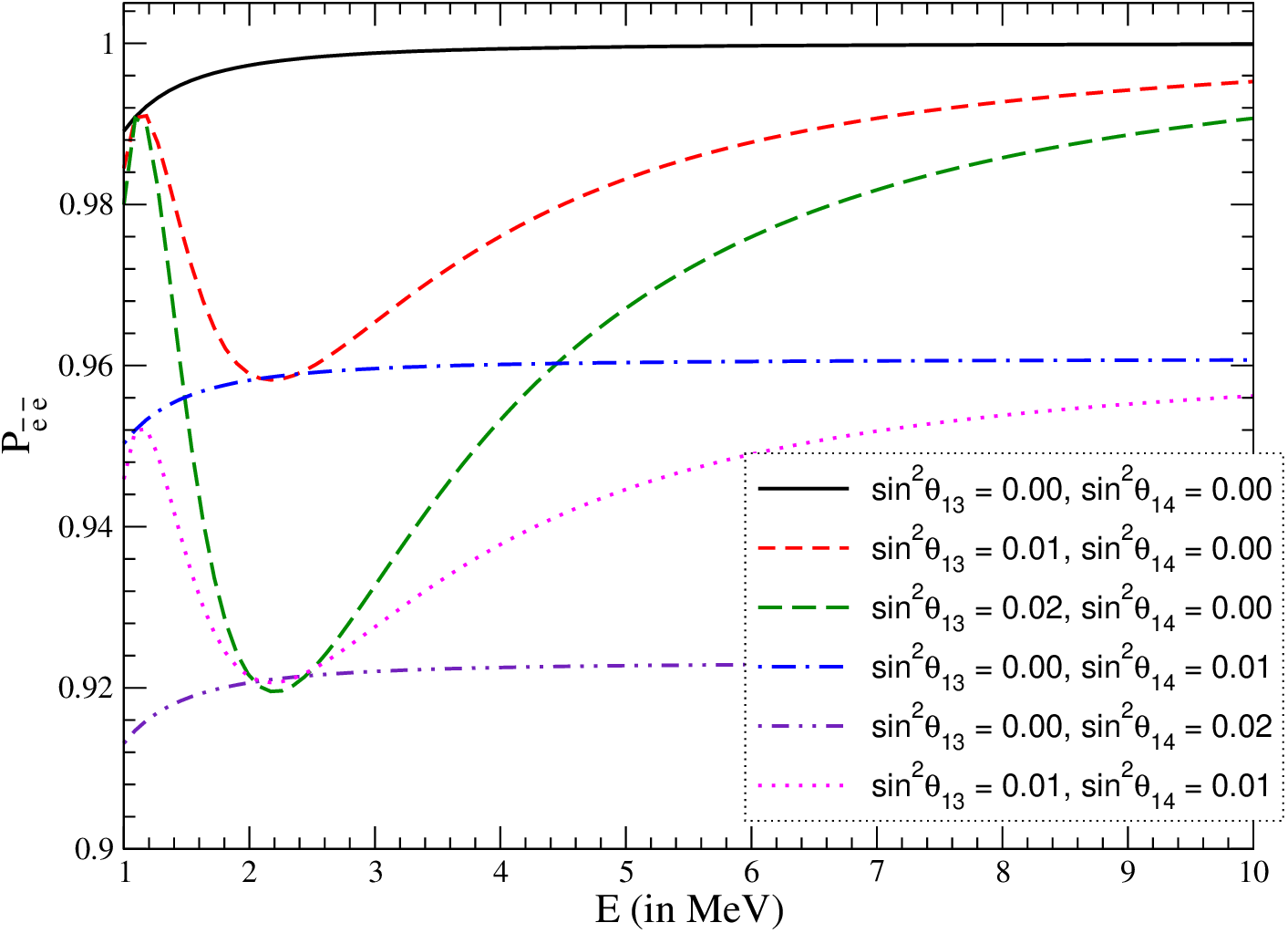}
\caption{\label{fig:pee}
The $\anue$ survival probability as a function of 
energy for the near (left panel) and far (right panel)
detectors of Double Chooz. The 6 different line types are 
for 6 combinations of $\sch$ and $\ses$ which are 
shown in the panels. We assume that 
$\sess=\ses$. 
}
\end{figure}

The probability channel relevant for Double Chooz is $P_{\bar e \bar e}$
which for the 3+2 mass spectrum is given by\footnote{Note that 
for the Double Chooz experiment there are neither any matter effects
nor any CP violating effect due the phases. 
The survival probability is therefore 
the same for neutrinos and antineutrinos.} 
\be
P_{\bar e\bar e} = 
1 - 4\sum_{i>j}|U_{ei}^s|^2|U_{ej}^s|^2\sin^2\left(\frac{\Delta m
_{ij}^2L}{4E} \right)
- 4\sum_{k>l}|U_{ek}^s|^2|U_{el}^s|^2\sin^2\left(\frac{\Delta m
_{kl}^2L}{4E}\right)~,
\ee
where $U^s$ is the mixing matrix and 
the indices $i$ and $j$ run from 1-3, $l$ runs from 1-5 and 
$k_s$ could be either 4 or 5. 
The 
oscillatory part of the third term of course will average 
out to 1/2 for Double Chooz, reducing the probability to
\be
P_{\bar e\bar e} = 
1 - 4\sum_{i>j}|U_{ei}^s|^2|U_{ej}^s|^2\sin^2\left(\frac{\Delta m
_{ij}^2L}{4E}\right) - 2\sum_{k>l}|U_{ek}^s|^2|U_{el}^s|^2~.
\ee
For the 3+2 mass spectrum we would have a $5\times 5$ mixing 
matrix for which we choose the convention 
\be
U^s = R(\theta_{45})R(\theta_{35})R(\theta_{34})R(\theta_{25})
R(\theta_{24})R(\theta_{15})R(\theta_{14})
R(\theta_{23})R(\theta_{13})R(\theta_{12})~,
\label{eq:uspara}
\ee
where $R(\theta_{ij})$ are the rotation matrices 
and $\theta_{ij}$ the mixing angle. We do not show the 
CP phases in Eq. (\ref{eq:uspara}) for simplicity. 
The mixing matrix with the above convention is
expressed as 
\be
U^s = \pmatrix
{c_{15}c_{14}c_{13}c_{12} & c_{15}c_{14}c_{13}s_{12}
& c_{15}c_{14}s_{13} & c_{15}s_{14}
& s_{15} \cr
. & . & . & . & \cr
. & . & . & . & \cr
. & . & . & . & \cr
. & . & . & . & \cr
}
~,
\label{eq:us}
\ee
where $c_{ij}=\cos\theta_{ij}$, $s_{ij}=\sin\theta_{ij}$, and 
we show explicitly only the first row since the 
probability $P_{\bar e \bar e}$, 
involves only them.
We can see that apart from the two usual mixing angles
$\theta_{12}$ and $\theta_{13}$ which appear in $P_{\bar e\bar e}$ 
with standard three generation oscillations, we 
have 2 additional angles, $\theta_{14}$ and $\theta_{15}$, 
which will affect the probability. If these mixing angles were 
zero, one would get back the $P_{\bar e \bar e}$ 
predicted by three generation oscillations. Of course 
for the energy and baseline of 
Double Chooz, the oscillations due to $\ms$ are extremely
weak and as a result so is the dependence on $\theta_{12}$. 
Therefore the dominant dependence of the probability would be 
on the mass squared difference $\ma$ and the mixing angles 
$\theta_{13}$, $\theta_{14}$ and $\theta_{15}$.

Current $3\sigma$ 
constraints \cite{limits} on the parameters driving the 
leading oscillations in solar \cite{solar}, atmospheric \cite{atm},
K2K \cite{k2k}, MINOS \cite{minos} and KamLAND \cite{kl} are 
\be
7.2\times 10^{-5} {\rm eV}^2 < \ms < 9.2\times 10^{-5} {\rm eV}^2 ~,
\label{eq:ms}
\ee
\be
0.25 < \sss < 0.39 ~,
\label{eq:sss}
\ee
\be
2.0\times 10^{-3} {\rm eV}^2 < \ma < 3.2\times 10^{-3} {\rm eV}^2  ~,
\ee
\be
\sin^22\theta_{23} > 0.9 ~.
\ee
The best limit on the mixing angle $\theta_{13}$ comes from the  
combined constraints from global oscillation data
including CHOOZ \cite{chooz} and is given as 
\cite{limits}
\be
\sch < 0.044~.
\label{eq:sch}
\ee
The sterile sector receives constraints from the short baseline 
reactor and accelerator based experiments \cite{sbl} which reported 
null signal, the LSND experiment \cite{lsnd} and MiniBooNE \cite{miniboone}.
The best-fit values for the mass squared difference $\Delta m^2_{41}$ 
and $\Delta m^2_{51}$ are 0.87 eV$^2$ and 1.91 eV$^2$ respectively, 
if the low energy MiniBooNE data is also included \cite{maltonischwetz}.
The best-fit values for the 
elements $U_{e4}$ and $U_{e5}$ of the mixing matrix are
0.12 and 0.11 respectively \cite{maltonischwetz}. Since in the 
convention adopted in this paper $U_{e5}=\sin\theta_{15}$
and $U_{e4} = \cos\theta_{15}\sin\theta_{14}$, this 
would translate to the best-fit values for the mixing angles as
$\sin^2\theta_{14}=0.012$ and $\sin^2\theta_{15}=0.015$. 

The Double Chooz experiment is being built to probe the mixing 
angle $\theta_{13}$. However, since the probability depends also 
on the sterile mixing angles $\theta_{14}$ and $\theta_{15}$, 
these angles also can be constrained in this experiment. The 
parameters $\Delta m^2_{41}$  and $\Delta m^2_{51}$ are of course 
averaged out and hence cannot be probed, and as discussed before,
the solar parameters $\ms$ and $\sss$ bring in a weak effect on 
$P_{\bar e \bar e}$.  
In what follows, 
we will keep $\ms$ and $\sss$ fixed at their best-fit values 
given in Eqs. (\ref{eq:ms}) and (\ref{eq:sss}). The true value of 
$\ma$ will be assumed to be $2.5\times 10^{-3}$ eV$^2$ throughout.
Also, {\it just for the sake of simplicity} we will take 
$\sin^2\theta_{14}=\sin^2\theta_{15}$ everywhere. 

In Fig. \ref{fig:pee} we show the probability $P_{\bar e \bar e}$ 
as a function of the antineutrino energy for 6 different choices 
of the mixing angles $\theta_{13}$ and $\theta_{14}$. For all 
other oscillation parameters we stick to the assumptions mentioned 
in the previous paragraph. The left panel of the figure shows 
$P_{\bar e \bar e}$ at a distance of 280 m, which will be 
the average distance of 
the near detector 
from the 2 cores of the Chooz-B
reactor. The right panel shows 
$P_{\bar e \bar e}$  for the far detector which will be at an  
average distance of 1.05 km 
from the 2 reactor cores. 
For 
the near detector, when 
$\theta_{14}=0=\theta_{15}$, we expect $\peebar$ 
to be almost 1 at high energies. However at lower energies,
for $E \ltap 4 $ MeV there is a $\ma$ driven dip in the 
survival probability, the extent of the dip depending on the 
value of $\sin^22\theta_{13}$. When we put $\theta_{13}=0$ and 
allow $\theta_{14}$ and $\theta_{15}$ to be non-zero, 
we get an energy independent suppression, depending on the 
value of these mixing angles. 
When all three mixing angles are non-zero 
we have an energy independent suppression driven by 
$\theta_{14}$ and $\theta_{15}$ superposed on the 
energy dependent dip at low energy due to $\theta_{13}$.  
At the far detector the $\ma$ driven oscillations of course are 
absolutely pronounced and the amplitude of the oscillations 
are determined by the value of $\sin^22\theta_{13}$. The 
energy independent average oscillation due to the extra large 
mass squared differences due to the sterile states are superimposed 
on the standard oscillations. We can see from the figure that 
more than one combination of $\theta_{13}$ and 
$\theta_{14}$ (and $\theta_{15}$) would 
give the same total suppression of the 
flux due to oscillations at the far detector. 
However, the shape of the resultant flux at the 
detector is expected to be different for the 
$\theta_{13}$ and $\theta_{14}$ (and $\theta_{15}$)
dependent oscillations.

\section{Results}

\subsection{Number of Events at the Near and Far Detector}

\begin{figure}
\includegraphics[width=8.0cm, height=7cm, angle=0]{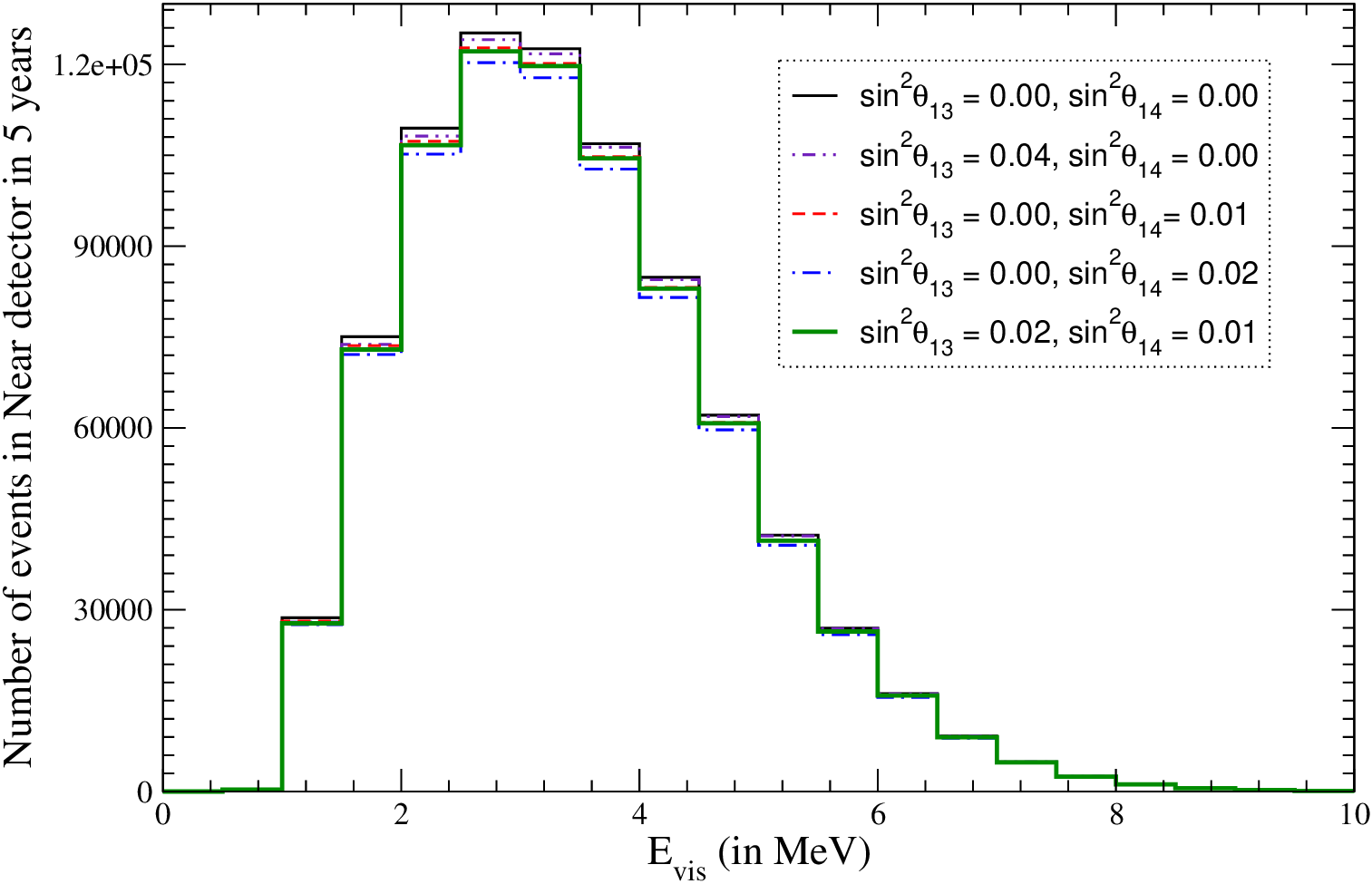}
\vglue -7.0cm \hglue 9.0cm
\includegraphics[width=8.0cm, height=7cm, angle=0]{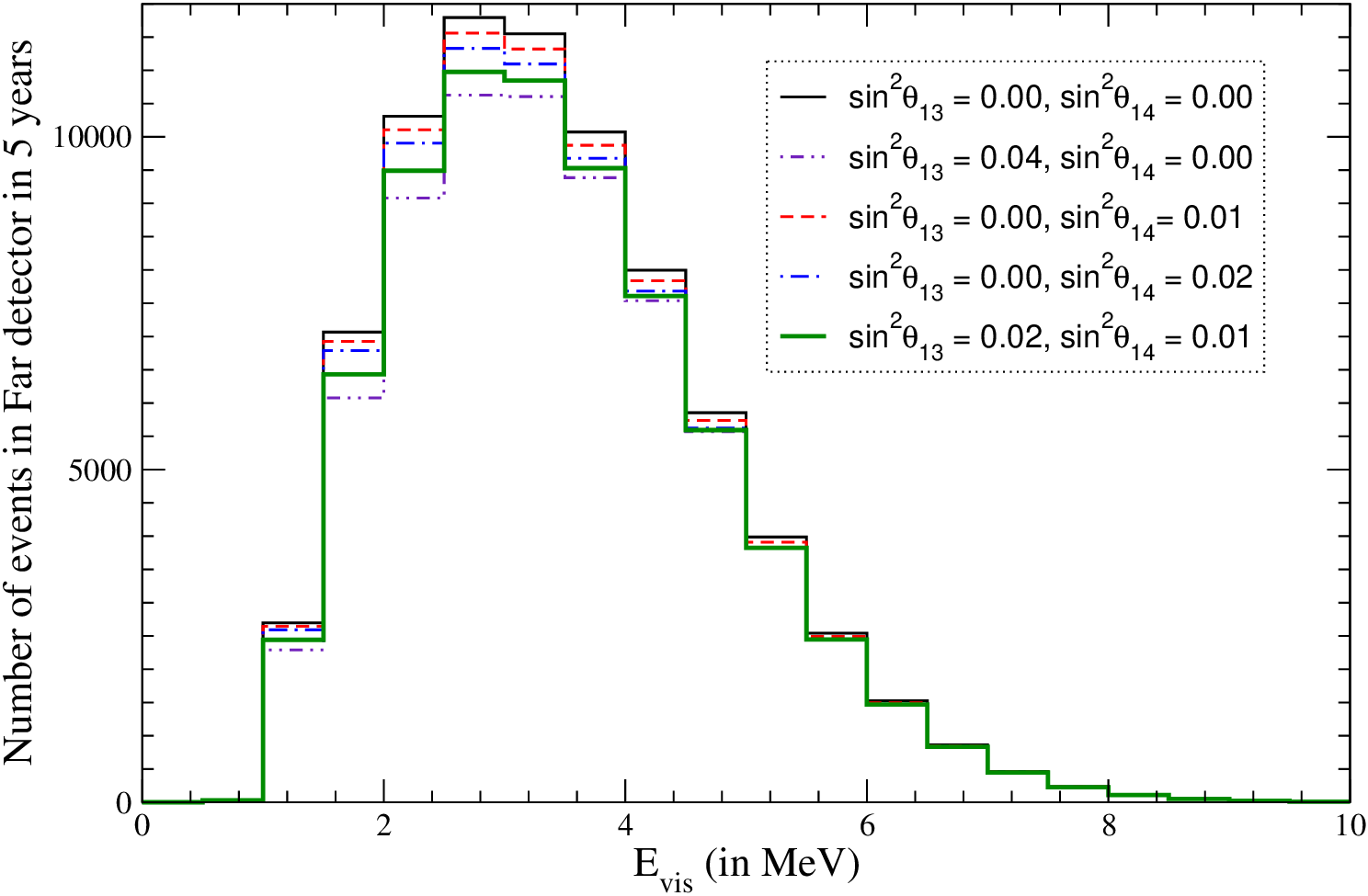}
\caption{\label{fig:rateE}
Number of events in 5 years expected at the near (left panel) and 
far (right panel) detectors of Double Chooz, as a function of the 
``visible energy'' $E_{vis}$ of the detected positron. 
 The 5 different line types are 
for 5 combinations of $\sch$ and $\ses$ which are 
shown in the panels. We assume that 
$\sess=\ses$. 
}
\end{figure}

\begin{figure}
\includegraphics[width=8.0cm, height=7cm, angle=0]{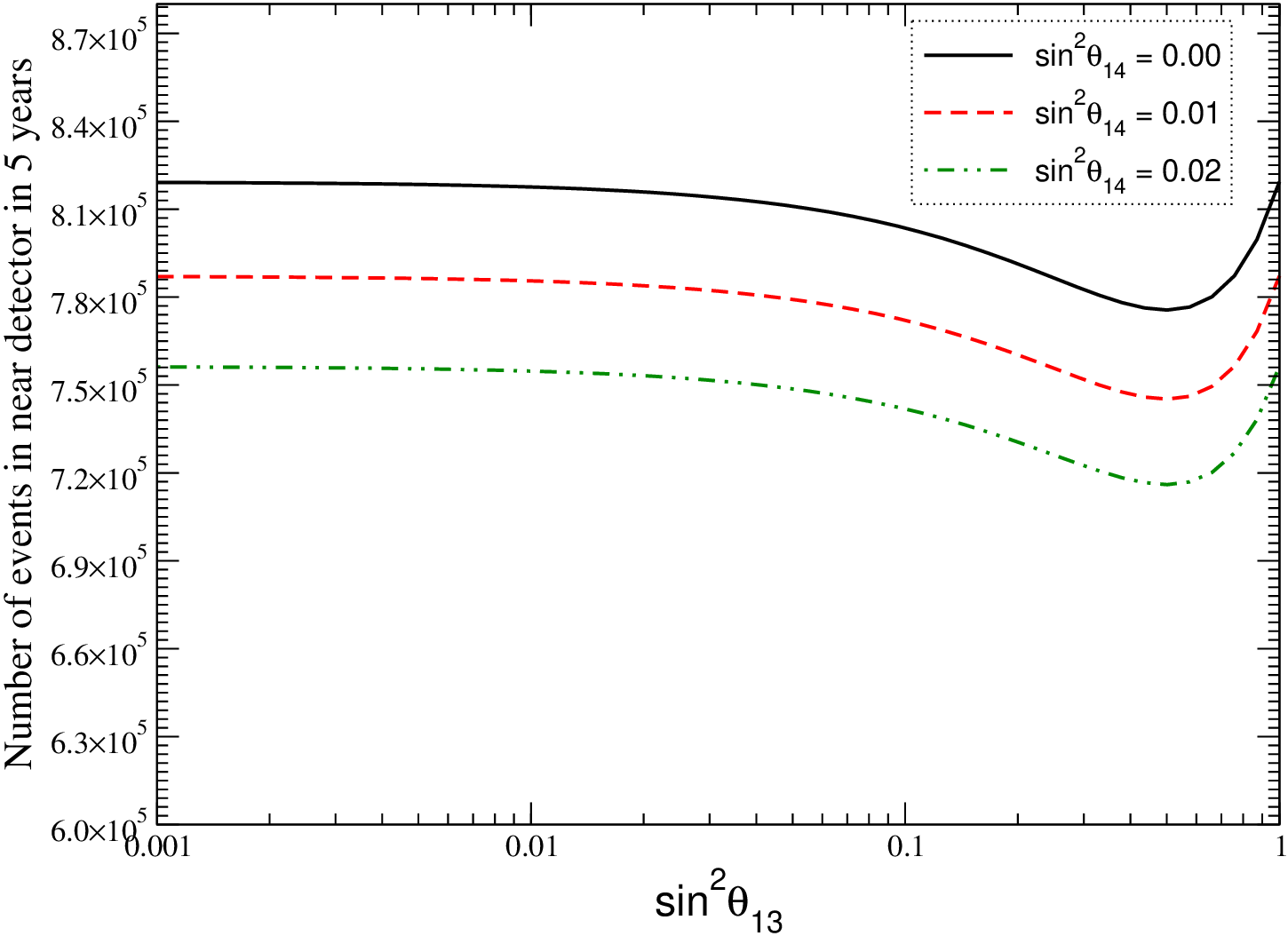}
\vglue -7.0cm \hglue 9.0cm
\includegraphics[width=8.0cm, height=7cm, angle=0]{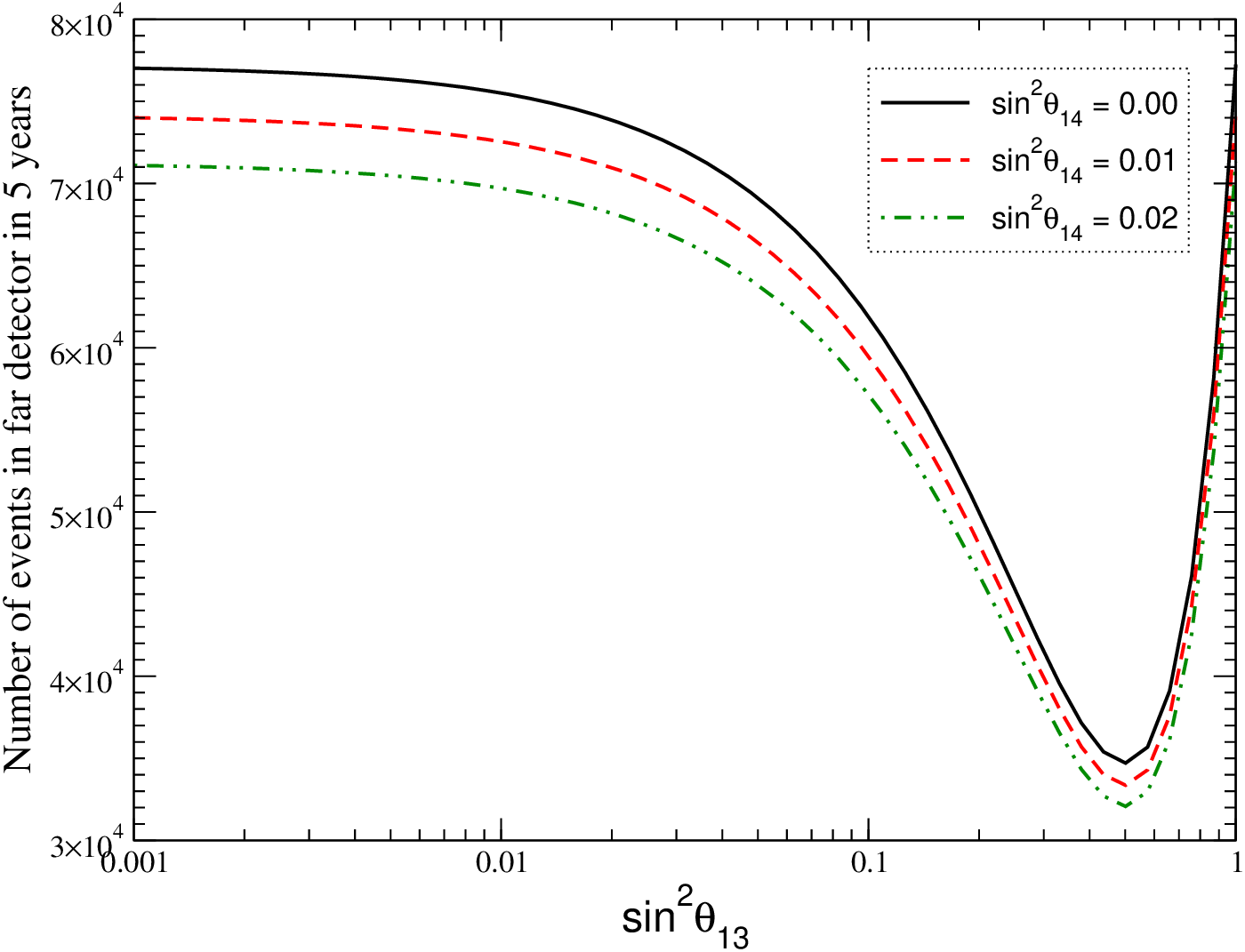}
\caption{\label{fig:rate13}
Total number of events in 5 years expected at the near (left panel) and 
far (right panel) detectors 
of Double Chooz, as a function of $\sch$. 
The black solid line is for $\ses=0$, red dashed for 
$\ses=0.01$ and green dot-dashed for $\ses=0.02$. 
   We assume that 
$\sess=\ses$. 
}
\end{figure}

The Chooz-B reactor complex consists of two reactor cores,
east ($R_E$) and west ($R_W$), 
with a total thermal power of 4.27 GW each. The $\anue$ 
flux at the near ($N$) and far ($F$) 
detector can be written as
\be
\Phi(E) = \frac{1}{4\pi L^2_{N,F}}\sum_{i}N_{i}^{fis} \phi_i(E)~,
\ee
where $L_{N,F}$ is the distance from the reactor core to the 
near ($N$) or far ($F$) detector\footnote{The far detector 
is at a distance 1114.6 m and 997.9 m from $R_E$ and $R_W$ 
respectively,
while the near detector is 290.7 m and 260.3 m 
away from them.},
$ N_{i}^{fis} $ are the number of fissions per second for the 
isotope $i$ in the reactor which we take from  \cite{Bemporad:2001qy}
and $\phi_i(E)$ gives the 
corresponding energy spectrum 
\begin{eqnarray}
\phi_i(E) = \exp\left(\sum_{k=0}^2 a_{ki}E^k\right)~.
\end{eqnarray}
We assume a second order polynomial parameterization 
of $a_{ki}$ 
for the four
isotopes 
($i =$ $^{235}U$, $^{239}Pu$, $^{241}Pu$, $^{238}U$).
The coefficients are taken from Table 2 of \cite{Huber:2004xh}.
The $\anue$ are 
detected through their capture on protons
\be
\anue + p \rightarrow e^+ + n
\ee
whereby the $e^+$ and $n$ give the prompt 
and delayed signal respectively in coincidence. 
The near and far detectors will be almost identical, 
consisting of a target volume of 
10.32 $m^3$ of liquid scintillator comprising 
of 80\% dodecane and 20\% PXE, leading to $6.79\times 10^{29}$ 
free target protons. The scintillator is doped with 
0.1\% gadolinium. 
The computed number of positron events in $n^{th}$ 
energy bin in the detector is given by
\begin{eqnarray}
N_{N,F}^{n} = 
F_{R} \int_{E_n}^{E_{n+1}} dE_{vis}
\int_0^\infty dE\ 
\sigma(E)\ R(E,E_{vis})\ P_{\bar{e}\bar{e}}(E,L_{N,F})\ 
\sum_i \frac{N_i^{fis} \phi_i(E)}{4\pi L_{N,F}^2}~,
\label{eq:ev}
\end{eqnarray}
where $E_{vis}$ is the measured {\it visible} 
energy of the emitted positron 
when the true visible energy $E_{vis}^T\simeq E - 0.8$ MeV, 
with $E$ being the energy of the incoming reactor antineutrino. 
The reaction cross-section is denoted by $\sigma(E)$ and 
$R (E,E_{vis})$ is the energy resolution function of the 
detector. 
The quantity $F_R$ is given by
\begin{eqnarray}
F_{R} &=& G \times P \times N_p\times T \times (1-d_D) \times \epsilon_D~,
\end{eqnarray}
where,
$G$ is the Global Load factor (reactor efficiency),
$P$ is the Reactor thermal power,
$N_p$ are the number of protons in the target volume,
$T$ is the exposure time,
$d_D$ is the dead time fraction of Detector D and
$\epsilon_D$ is the detector efficiency.
Values of all quantities needed for calculating the 
positron rate in the detector 
are taken from \cite{chooz2}.

We show in Fig. \ref{fig:rateE} the number of events expected 
in 5 years 
as a function of the visible positron 
energy, for the near (left panel)
and far detector (right panel). 
We show the event spectra for 
5 different combinations of $\theta_{13}$ and 
$\theta_{14}$ (and $\theta_{15}$) values. For the near 
detector it is mainly $\theta_{14}$ (and $\theta_{15}$) values
which cause difference to the event spectra. At the far detector 
oscillations with both frequencies 
are important and hence all the three mixing angles 
make a difference. 
We stress that even though 
we do not show the statistical errorbars on this figure for the 
sake of clarity, one can easily check that most of the 
cases of mixing parameters displayed on this figure should be 
statistically distinguishable by combining the near and far 
event spectra at Double Chooz.

In Fig. \ref{fig:rate13} we show the total observed positron 
events in 5 years as a function of $\sch$, 
for the near (left panel)
and far detector (right panel). 
The three different line types in either of the panels 
show the results for a different choice for the value of 
$\theta_{14}$ (and $\theta_{15}$). The net suppression of course 
increases as $\sin^2\theta_{14}$ increases. This is true at both the 
near and far detectors. However, the $\sch$ dependence of the 
suppression is  
extremely mild at the near detector and very large at the far detector.
As expected, the largest suppression comes at $\sch=0.5$ at which 
we have maximal oscillations. The most important thing we can note 
from this figure is that for a certain range of events at 
the far detector, 
any given observed rate could be predicted by 
a wide set of possible values of 
$\theta_{13}$ and $\theta_{14}$ (and $\theta_{15}$). 
All these would then constitute degenerate solutions. 
For instance, we can see from the figure that 
if the far detector was to observed 
$7\times 10^4$ events in 5 years, then this would allow 
the sets of ($\theta_{13}$, $\theta_{14}$) values, 
(0.043, 0.00), (0.026, 0.01) and (0.008, 0.02), 
as possible solutions. These would be degenerate solutions 
in this case. However, we can see by looking at the left panel that  
each of these combinations would predict different total rates at 
the near detector. Therefore by combining the near and far detected 
event rates, one can overcome this degeneracy problem.

\subsection{The $\theta_{13}$ sensitivity}

\begin{figure}
\includegraphics[width=8.0cm, height=7cm, angle=0]{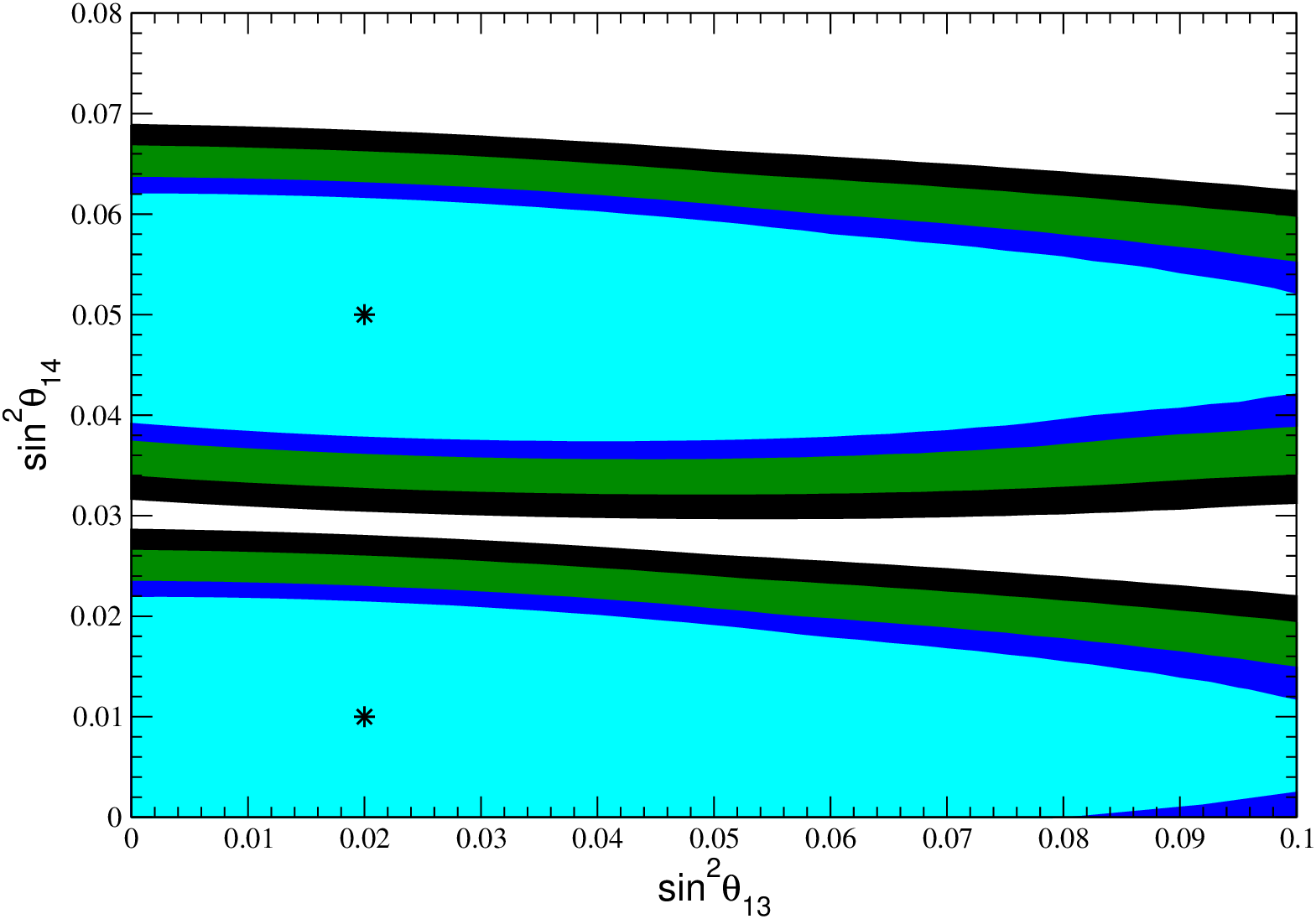}
\vglue -7.0cm \hglue 9.0cm
\includegraphics[width=8.0cm, height=7cm, angle=0]{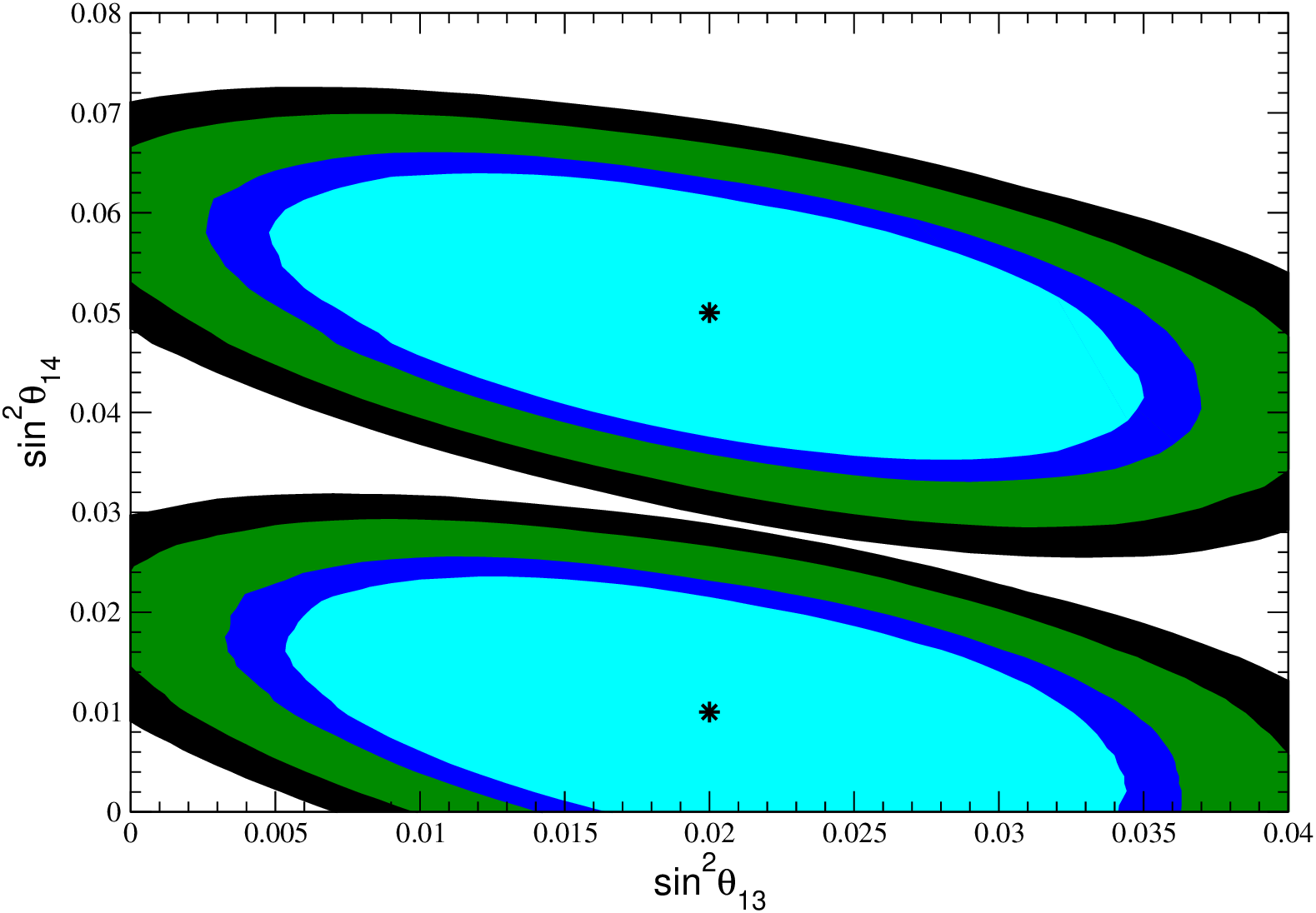}
\caption{\label{fig:cont}
The 90\%, 95\%, 99\% and 99.73\% C.L. contours in the 
$\sch-\ses$ plane for different assumed true values of the 
mixing angles marked in the figure by stars. 
Left panel is for the data from near detector only while the 
right panel is for data from far detector alone. 
 We assume that 
$\sess=\ses$. 
}
\end{figure}

\begin{figure}
\begin{center}
\includegraphics[width=12.0cm, height=7cm, angle=0]{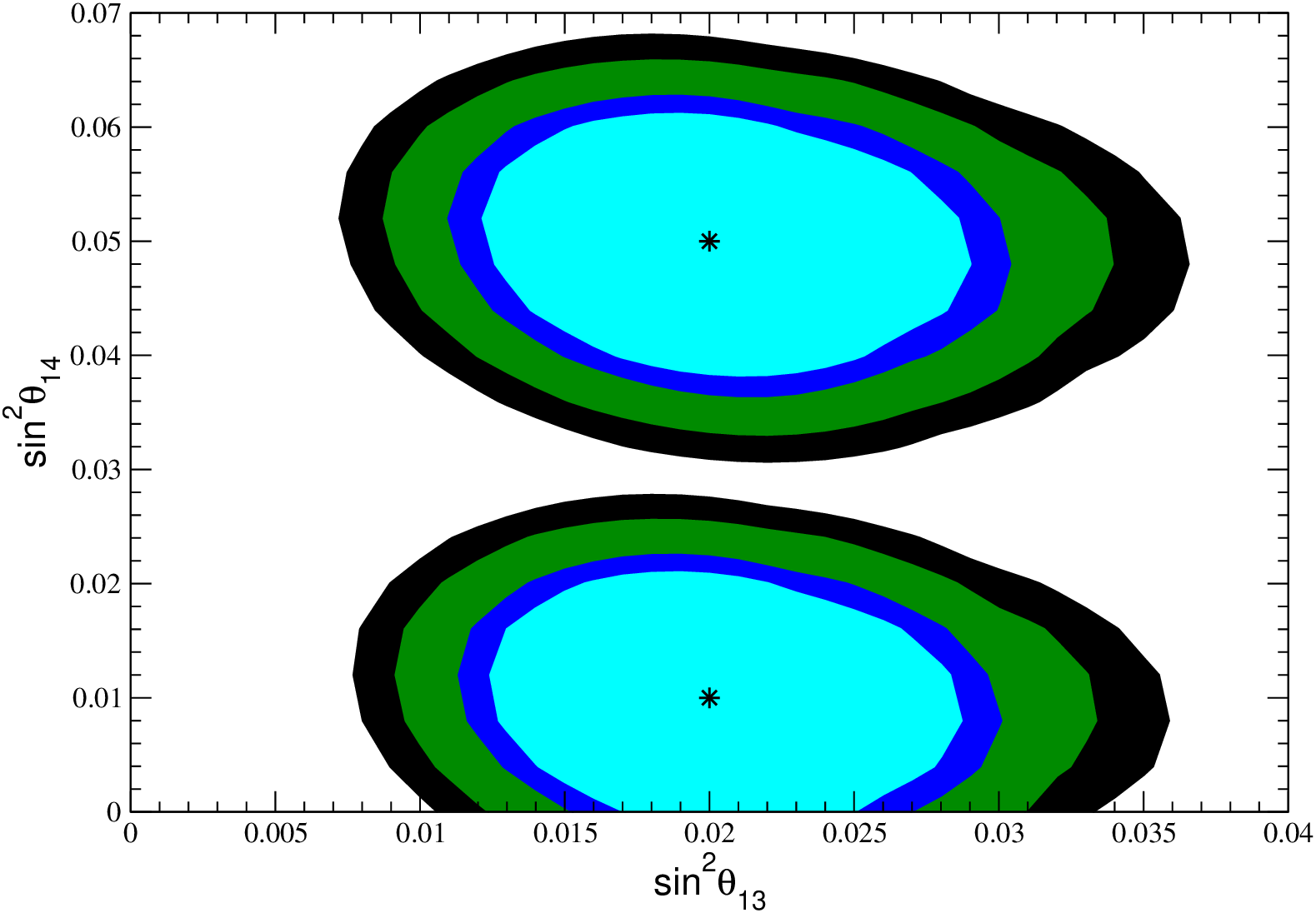}
\caption{\label{fig:contNF}
The 90\%, 95\%, 99\% and 99.73\% C.L. contours in the 
$\sch-\ses$ plane for assumed true values of the 
mixing angles marked in the figure by stars. 
Data from both near and far detectors are combined. 
We assume that 
$\sess=\ses$. 
}
\end{center}
\end{figure}
\begin{figure}
\begin{center}
\includegraphics[width=17.0cm, height=7cm, angle=0]{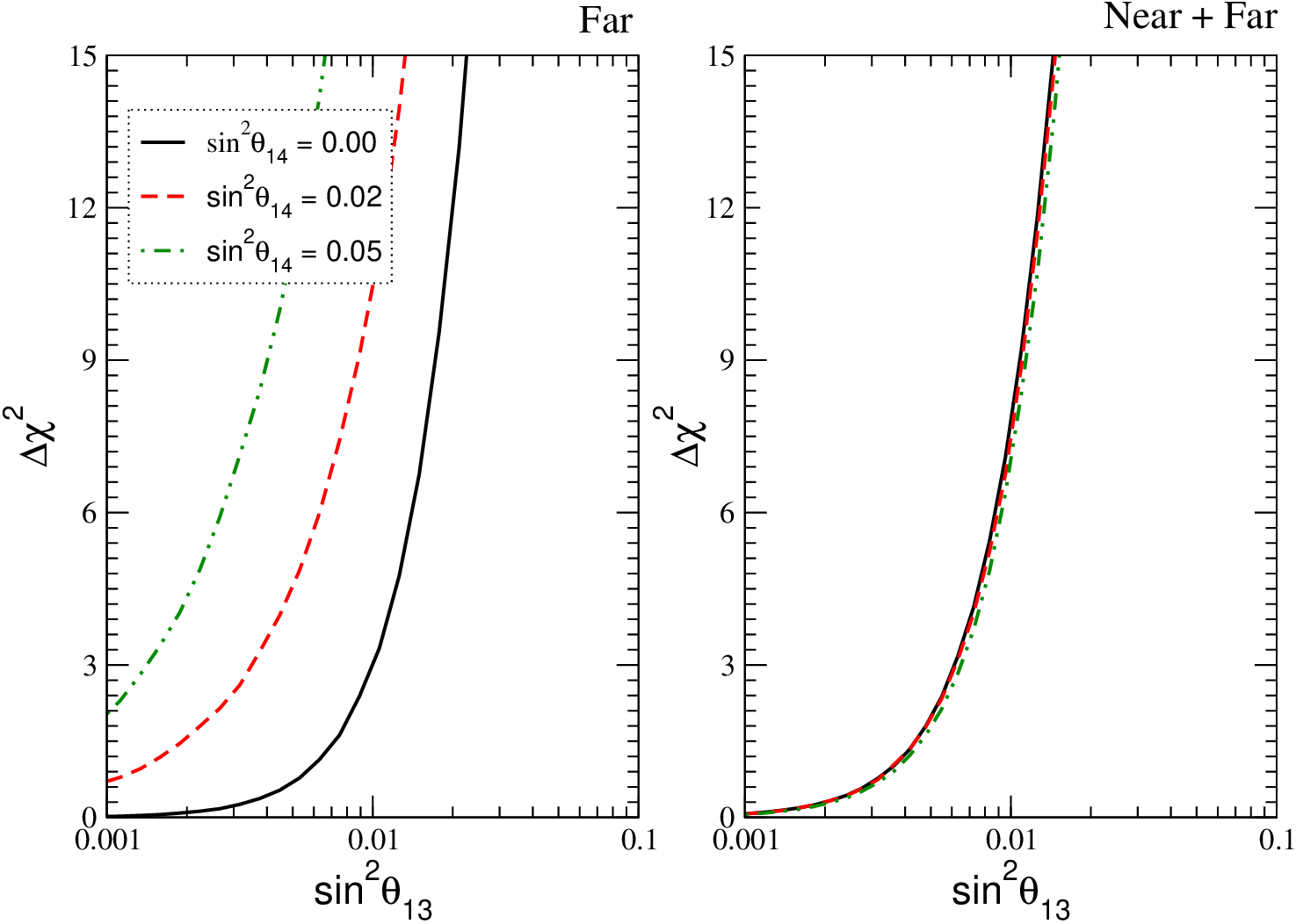}
\caption{\label{fig:sen}
Sensitivity plot showing the $\Delta \chi^2$ as a function of 
$\sch$ for the far detector only (left panel) and when 
near and far data sets are combined (right panel). Data is 
generated at $\sch=0$ and $\ses=0$ for all curves in both panels.
In the fit we fix $\ses=0$ (black solid line), 0.02 (red dashed line)
and 0.05 (green dot-dashed line). 
We assume that 
$\sess=\ses$. 
}
\end{center}
\end{figure}

\begin{figure}
\begin{center}
\includegraphics[width=17.0cm, height=7cm, angle=0]{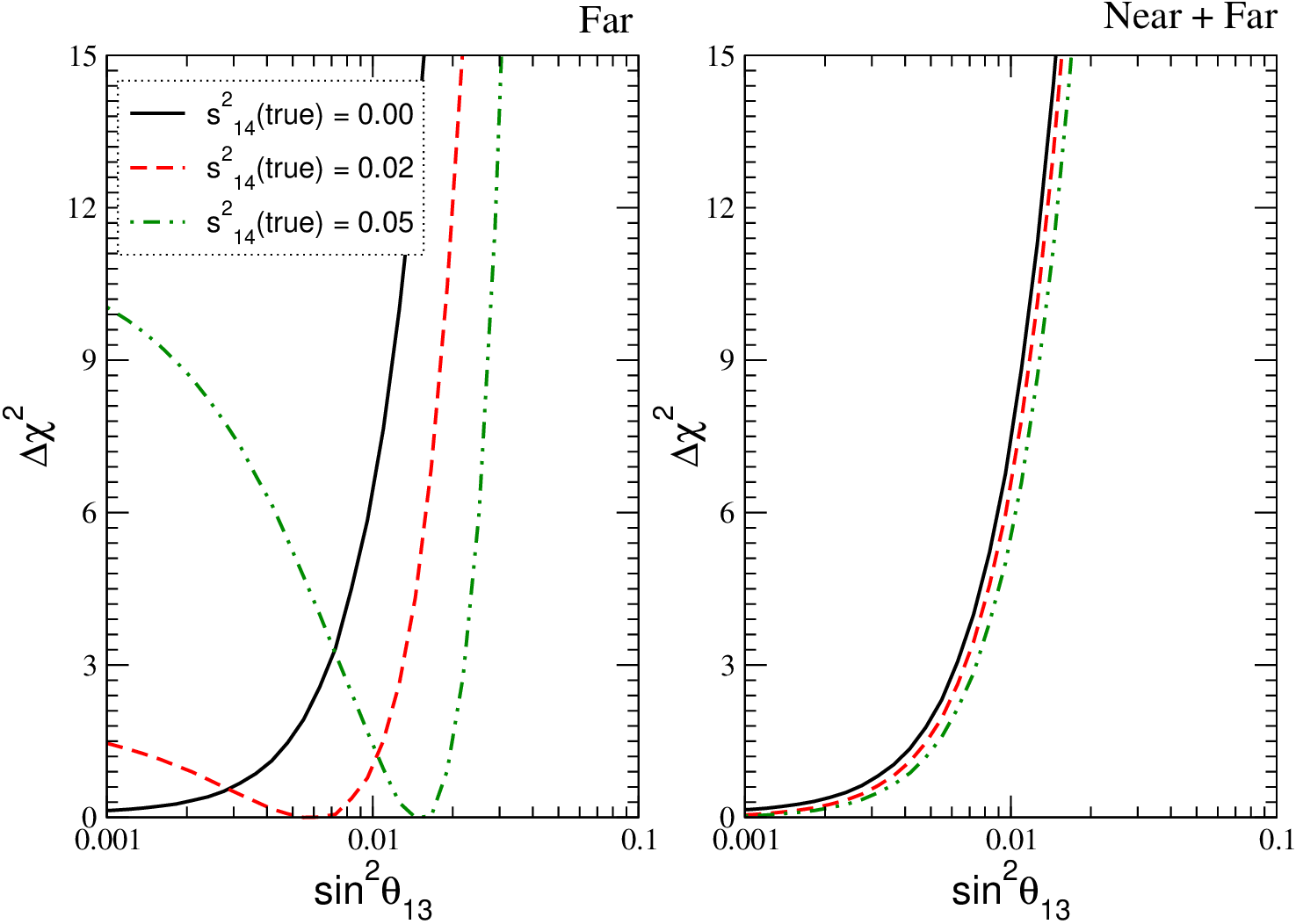}
\caption{\label{fig:sen2}
Sensitivity plot showing the $\Delta \chi^2$ as a function of 
$\sch$ for the far detector only (left panel) and when 
near and far data sets are combined (right panel). 
Data is generated at $\sch=0$ and $\ses=0$ (black solid line)
0.02 (red dashed line) and 0.05 (green dot-dashed line). 
In the fit we fix $\ses=0$ for all curves.
We assume that 
$\sess=\ses$. 
}
\end{center}
\end{figure}

\begin{figure}
\begin{center}
\includegraphics[width=17.0cm, height=7cm, angle=0]{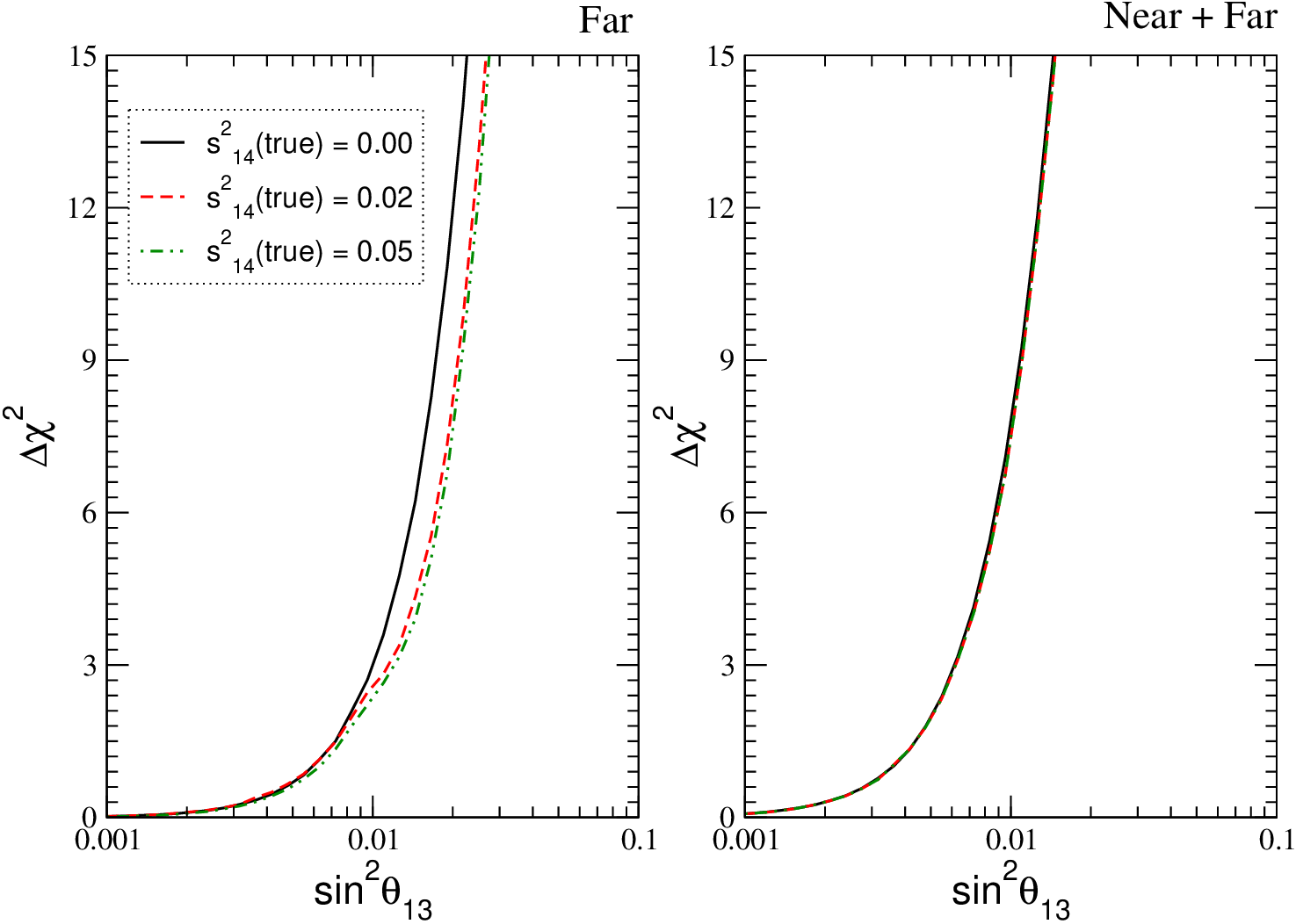}
\caption{\label{fig:sen3}
Sensitivity plot showing the $\Delta \chi^2$ as a function of 
$\sch$ for the far detector only (left panel) and when 
near and far data sets are combined (right panel). 
Data is generated at $\sch=0$ and $\ses=0$ (black solid line)
0.02 (red dashed line) and 0.05 (green dot-dashed line). 
In the fit $\ses$ is allowed to vary freely and take any possible 
value for all curves.
We assume that 
$\sess=\ses$. 
}
\end{center}
\end{figure}

For estimating the projected sensitivity of the Double Chooz 
experiment to the mixing angles $\theta_{13}$, $\theta_{14}$ 
and $\theta_{15}$, we perform statistical analysis of 
5 years prospective data at both the near and far detectors. 
We define a $\chi^2$ function on the lines of 
\cite{chooz2}
\begin{eqnarray}
\chi^2_{R}(\Theta_{true}) 
&=& \sum_{D=N,F} \sum_i
\frac{[(1+a_{corr}+a^D_{uncorr}+a^i_{spect})N^i_{D}(\Theta_{fit}) 
- N^i_D(\Theta_{true})]^2}{N^i_{D} + B^i_D + (N^i_{D}\sigma_{bin})^2+
(B^i_{D}\sigma_{bkd})^2}
\nonumber \\
&& + \frac{a_{corr}^2}{\sigma_{corr}^2}
+ \sum_{D=N,F} \frac{(a^D_{uncorr})^2}{(\sigma_{uncorr})^2}
+ \sum_{i} \frac{(a^i_{spect})^2}{(\sigma_{spect})^2}~,
\label{eq:chisq}
\end{eqnarray}
where $D$ runs for the number of detectors (near and far)
and $i$ runs over the number of bins, 
$\Theta_{true}$ are the set of oscillation parameters
at which the data is generated and $\Theta_{fit}$ is the 
corresponding set in theory. The systematic errors taken 
into account can be 
broadly characterized as normalization errors
and spectral shape errors. The normalization errors 
include the 2\% overall reactor antineutrino 
flux uncertainty which is relevant for both 
the detectors and therefore correlated. We denote this by
$\sigma_{corr}$. The reminiscent uncorrelated normalization 
error between the detectors is taken as 0.6\% and is
denoted as $\sigma_{uncorr}$. The uncertainty in the 
reactor $\anue$ spectral flux is denoted by 
$\sigma_{spect}$ and we take 2\% as its estimated value.
This error is totally uncorrelated between the near and far detectors.
There is also a bin-to-bin uncorrelated systematic error which is 
denoted by $\sigma_{bin}$ and taken as 1\%. 
We also include the background subtraction error
$\sigma_{bkd}$ which is taken as 1\% of the background, $B^i_D$, 
and we assume that there is 1\% background. We have checked 
that the background and 
its corresponding error makes almost no 
difference to our final results. 
For each set of oscillation parameter value 
taken in the fit, 
the function $\chi^2_R(\Theta_{true})$ 
is minimized with respect to the parameters 
$a_{corr}$, $a_{uncorr}^D$ and $a_{spect}^i$, which are 
allowed to vary freely. 

In Eq. (\ref{eq:chisq}) we have explicitly put the subscript 
$R$ on the $\chi^2$ function to denote the contribution 
coming from the Double Chooz set-up alone. In our numerical 
analysis, we have also included a ``prior'' on the allowed 
values of $\Delta m_{31}^2$ since we expect that the 
uncertainty on this parameter will see some reduction by the 
time the Double Chooz results are declared. Our full $\chi^2$ 
is therefore given by
\be
\chi^2 = \chi^2_R + \chi^2_{prior}~,
\label{eq:chitot}
\ee
where,
\be
\chi^2_{prior} = \bigg(\frac{\ma - \Delta m_{31}^2(true)}
{\sigma_{\ma}}\bigg)^2~,
\ee
where we assume that 
$\Delta m_{31}^2(true)=2.5\times 10^{-3}$ eV$^2$ 
and $\sigma_{\ma}$ is 10\% of 
$\Delta m_{31}^2(true)$.
The total $\chi^2_{tot}$ is then minimized with respect to 
some or all the oscillation parameters, to 
obtain the best-fit values, sensitivity 
limits and C.L. contours. In all our results presented in this 
paper, we marginalize over $\ma$. We 
keep $\ms$ and $\sss$ fixed at their best-fit value due 
to reasons  discussed
before. For the mixing angles $\theta_{13}$, $\theta_{14}$ and 
$\theta_{15}$, we will always mention whether they are free or
fixed in the fit.

In Fig. \ref{fig:cont} we show the expected C.L. contours in the 
$\sch-\ses$ plane. 
The left panel shows contours expected from 
analysis of data from the near detector only, while right panel
shows the corresponding results when only the far detector data 
is analyzed. The points at which the data were generated are shown 
by the ``star'' marks in the figure. The different color shades 
show the 90\%, 95\%, 99\% and 99.73\% contours.
The $\Delta \chi^2$ for the C.L. contours correspond to 2 parameters.  
We see that the far detector can simultaneously constrain 
$\theta_{13}$ and the sterile mixing angles $\theta_{14}$ and 
$\theta_{15}$. If the true value of $\sch=0.02$ and $\ses=0.05$, 
we could measure the sterile mixing angles within the range,
$0.025 \leq \ses \leq 0.072$,
at $3\sigma$. If the true value was 
$\sch=0.02$ and $\ses=0.01$ 
the lower limit for 
$\ses$ would be restricted by 0 from below so that we would have
$0.0 \leq \ses \leq 0.03$. 
For $\sch$ we see that at 95\% C.L.\footnote{We give the 
95\% C.L. for $\sch$ since above this the contours cross 
the y-axis.}
$0.0026 \leq \sch \leq 0.037$ when data is at 
$\sch=0.02$ and $\ses=0.05$, and 
$0.003 \leq \sch \leq 0.036$ when data is at 
$\sch=0.02$ and $\ses=0.01$.
We note from this that the limits 
on $\sch$ at the far detector depends on the 
value of $\ses$, albeit very slightly. Indeed the 
tilt of the C.L. contours towards the left shows a mild 
anticorrelation between the two mixing angles. We remind the 
reader that we keep $\sess=\ses$ fixed
throughout this paper. The contours for the near detector 
explicitly show that there is hardly any sensitivity to $\sch$, 
unless its value was very large, which anyway is already 
disfavored. It could however restrict at $3\sigma$ the sterile 
mixing 
to $0.03 \leq \ses \leq 0.069$ if the true value was 
$\ses=0.05$ and $0.0 \leq \ses \leq 0.028$ if the true value was 
$\ses=0.01$. We see that the sterile mixing angle can be determined 
pretty well in either the near or far detector, with the precision 
in near detector being better due to its larger statistics. 

In Fig. \ref{fig:contNF} we show the allowed areas at 
90\%, 95\%, 99\% and 99.73\% C.L. for 2 parameter fit
when data from near and far detector are combined using the 
full expression given by Eq. (\ref{eq:chisq}). 
For the case where the data was simulated at 
$\sch=0.02$ and $\ses=0.05$, the $3\sigma$ limits on the 
mixing angles are 
$0.007 \leq \sch \leq 0.037$ and $0.031 \leq \ses \leq 0.068$.
For data at $\sch=0.02$ and $\ses=0.01$, corresponding 
limits are 
$0.008 \leq \sch \leq 0.036$ and $0.0 \leq \ses \leq 0.028$
We stress that we have given here the $3\sigma$ limits
for $\sch$ while in the previous paragraph they 
were at the 95\% C.L. 
Note that the sensitivity to $\sch$ improves dramatically.
The reason of course is that the near-far detector combination 
effectively assures that the correlated systematic errors 
cancel out, leaving just the small uncorrelated part. 
The uncertainty on $\ses$ and $\sess$ is also somewhat reduced. 
Another significant feature we notice from the 
figure is that the near-far detector combination breaks the 
anticorrelation between $\sch$ and $\ses$ seen for the 
far detector in Fig. \ref{fig:cont}.
The reason for this was discussed in the 
previous section.

If the true value of $\theta_{13}$ turns out to be very small 
or even zero then we would not see a positive signal 
at Double Chooz, at least in the three generation picture. 
We could nonetheless use the data to 
put an upper limit on the value of $\theta_{13}$ at a given 
C.L. In what follows, we will give the ``$\sch$ sensitivity reach'' 
for Double Chooz. For this we generate the data at 
$\sch=0$ and fit it back allowing for non-zero value for 
$\sch$. The resultant $\sch$ sensitivity reach is shown 
in Fig. \ref{fig:sen}, \ref{fig:sen2} and \ref{fig:sen3}.

In Fig. \ref{fig:sen} 
we have generated the data assuming  
the true values of $\sch=0$ and $\ses=0$ (and $\sess=0$).
We find the $\Delta \chi^2=(\chi^2(\sch)-\chi^2_{min})$ 
for every value of $\sch$,   
keeping $\ses$ (and $\sess$) fixed at 0 (solid black line) 
0.02 (red dashed line) and 0.05 (green dot-dashed line) in the fit. 
We note that the $\Delta \chi^2$ increases sharply with 
$\ses$ for the far detector alone. However, once we combine
the data sets from near and far detectors, the effect of 
$\ses$ becomes very small. This is related to the fact discussed 
before that the combined near and far data sets are almost 
independent of the correlation between $\sch$ and $\ses$
and can measure both of them independently. In other words,
the overall suppression or normalization factor 
due to $\ses$ gets canceled out from the fit for $\sch$, 
when data from the near detector is added to the data 
from the far detector.  

In Fig. \ref{fig:sen2}
we generate the data assuming the true value of $\sch=0$
and taking 
$\ses=0$ (black solid line), 0.02 (red dashed line)
and 0.05 (green dot-dashed line). Here we find the 
$\Delta \chi^2$ 
for every value of $\sch$, 
keeping $\ses$ (and $\sess$) fixed at 0 in the fit 
for all cases. This situation might easily arise in practice
if nature has indeed chosen a non-zero value for the sterile 
mixing angles and we being ignorant of that, try to fit the data by  
assuming that there were no sterile neutrinos mixed with the 
active ones. For the case where true value of $\ses$ was indeed 
0, we have the standard three generation case and here we 
recover the projected $\sch$ sensitivity for Double Chooz. 
If only data from far detector was taken then we could put the 
limit 
$\sch < 0.0095$ ($\sin^22\theta_{13} < 0.038$) at 90\% C.L.
and $\sch < 0.017$ ($\sin^22\theta_{13} < 0.047$) at $3\sigma$. 
For combined data from near and far detectors we could 
restrict the angle to
$\sch < 0.006$ ($\sin^22\theta_{13} < 0.023$) at 90\% C.L.
and $\sch < 0.011$ ($\sin^22\theta_{13} < 0.043$) at $3\sigma$.
However, if the true value of $\ses$ was non-zero, we would have 
some difference in the sensitivity limit, 
if we were using data from the 
far detector alone. We would get a signal at the 
detector due to the sterile neutrinos 
and might confuse it with a signal due to $\sch$.
It would look like we have 
observed a non-zero value of $\sch$.
We note from Fig. \ref{fig:sen2} that 
the best-fit $\sch$ could be as 
large as $\sch=0.012$ (0.027) if the true value of $\ses$ was 
0.02 (0.05).
This is the fake solution that we discussed about 
in the previous section. 
Once we combine the data sets 
from both near and far detectors, the impact of 
the sterile mixing angle is negated to a large extent. 
However, for very large values of $\ses$ such as 0.05, 
we find that some residual confusion and the fake solution remains.
Also, the upper limit on $\sch$ at a given C.L. 
turns out to be different 
compared to the case where $\ses$ was indeed zero.

Fig. \ref{fig:sen3} is similar to Fig. \ref{fig:sen2} 
in most respect, except that here we allow $\ses$ 
(and $\sess$) to take all possible values in the fit.
This would be the most democratic approach whereby 
all possible mixing angles are accounted for in the fit 
and marginalized over, while 
putting restrictions on $\theta_{13}$. 
The effect of the sterile mixing angle 
on the allowed values of $\sch$ in this case is not 
dramatic. This is mainly due to the fact that 
$\ses$ and $\sess$ can now take non-zero values in the fit
to give lowest possible $\chi^2$, while in Fig. \ref{fig:sen2}
the only mixing angle we could fiddle with was $\sch$ 
since $\ses$ and $\sess$ were fixed at 0.
When data at only the far detector is taken,
the projected sensitivity on $\sch$ gets worse 
for true non-zero $\ses$, compared to when $\ses$ 
was truly zero in Nature.
But once data
from the near detector is added, the problem is 
solved and the standard projected sensitivity on $\sch$ 
is restored.

\section{Conclusions}

The recent MiniBooNE results, even though disagree with the 
LSND data, leave ample room for existence of extra sterile 
neutrinos. In particular, the 3+2 neutrino mass spectra with 
2 extra sterile neutrinos mixed with the active ones can 
satisfy the world neutrino data if one allows for CP violation. 
In this paper we probed the implications of the 3+2 
mass spectrum for the Double Chooz reactor experiment. 
We showed how the event spectra at the near and far detectors 
of the Double Chooz experiment change when we allow for 
sterile mixing. The oscillations driven by the extra sterile 
neutrinos would produce a constant suppression at both the near and 
far detectors of Double Chooz. This is in contrast to the $\ma$ 
driven and $\theta_{13}$ dependent oscillations which are almost absent 
in the near detector and only register their signal at the 
far detector. In the far detector, the extra oscillations due to 
sterile mixing would be superimposed over the normal 
flavor oscillation signal.  
In particular, we established that this sensitivity of the 
far detector to both $\theta_{13}$ and the sterile mixing angles
leads to a correlation between the 2 completely different kinds 
of mixing angles.

We defined a $\chi^2$ function for the statistical analysis of the 
data and presented our results. We showed that the presence of 
sterile mixing angles alter, albeit slightly, the 
precision with which $\theta_{13}$ could be determined in 
Double Chooz. The sterile mixing angle can be determined 
pretty well in either the near or far detector, with the precision 
in near detector being better due to its larger statistics. 
The combined near and far detector data sets would be able to 
determine both mixing angles, though the precision in $\theta_{13}$ 
goes down as the true value of the sterile mixing angle increases.
We also studied how the ``sensitivity to $\sch$'' changes in 
presence of sterile neutrinos. We defined the sensitivity reach 
as the maximum value of $\sch$ which would be able to fit the 
data at the chosen C.L., when its true value is exactly zero. 
If the true value of sterile mixing is indeed non-zero and 
we kept them fixed at zero in our fit, 
we would get significantly different results on the upper limit
of $\sch$, even after combining the results from both near and far 
detectors. This problem would stay if we allowed the sterile mixing 
to vary freely in the fit, but analyzed results from the 
far detector only. We finally showed that the problem could be 
fully negated only by combining results from both detectors and 
allowing the sterile mixings to take all possible values in the 
fit.

In conclusion, presence of sterile neutrinos would leave 
its imprint on the signal at both the near and far detectors 
of the Double Chooz experiment. 
If only the far detector signal was considered, 
there is a possibility 
of confusing these sterile neutrino driven oscillations with 
active ones. However, by taking results from both detectors and 
allowing for the possibility of sterile 
mixing angles, one can probe both $\theta_{13}$ and the 
sterile mixing angles correctly at Double Chooz.


\end{document}